\documentclass[conference]{IEEEtran}
\IEEEoverridecommandlockouts
\usepackage[T1]{fontenc}
\usepackage{amsmath,amssymb,amsthm}
\usepackage{newtxtext,newtxmath}
\usepackage{microtype}
\usepackage{cite}
\usepackage[hidelinks]{hyperref} 
\usepackage[capitalize]{cleveref}
\usepackage{graphicx}
\usepackage{float}
\usepackage{subfig}
\usepackage{textcomp}
\usepackage{xcolor}
\usepackage{csquotes}
\usepackage{multirow}
\usepackage{todonotes}
\usepackage{todonotes}
\usepackage{tikz}
\usepackage{siunitx}
\usetikzlibrary{shapes.geometric, arrows}
\usepackage{xcolor}
\usepackage{balance}
\usepackage{flushend}
\usepackage{algorithm}
\usepackage{algpseudocode}
\usepackage{amsmath}
\usepackage{booktabs}
\usepackage{tabularx}
\usepackage{makecell}
\usepackage{array}
\usepackage{multirow}
\usepackage{booktabs}
\usepackage{makecell}
\usepackage{multirow}
\usepackage{amsthm}
\newtheorem{theorem}{Theorem}
\usepackage{xspace}
\xspaceaddexceptions{)}
\usepackage{glossaries}
\usepackage{flushend}
\usepackage{array}
\usepackage{booktabs}

\newglossaryentry{vdd}{
	name={\ensuremath{V_{\text{DD}}}},
	description={supply voltage},
	symbol=Vdd,
	sort=vdd}
\newglossaryentry{vth}{
	name={\ensuremath{V_{\text{TH}}}},
	description={threshold voltage},
	symbol=Vth,
	sort=vth}
\newglossaryentry{gnd}{
	name={GND},
	description={El. ground},
	symbol=GND}

\newacronym{tcad}{TCAD}{technology CAD}

\newacronym{fefet}{FeFET}{ferroelectric FET}
\newacronym{fe}{FE}{ferroelectric}
\newacronym{ncfet}{NCFET}{negative capacitance FET}
\newacronym{feram}{FeRAM}{ferroelectric RAM}

\newacronym{fdsoi}{FD-SOI}{fully depleted silicon on insulator}
\newacronym{tcam}{TCAM}{Ternary Content Addressable Memory}

\newacronym{sap}{SAP}{stuck-at-polarization}

\newacronym{beol}{BEOL}{back-end-of-line}
\newacronym{feol}{FEOL}{front-end-of-line}
\newacronym{mfm}{MFM}{metal-ferroelectric-metal}

\newacronym{po}{PO}{primary output}
\newacronym{pi}{PI}{primary input}
\newacronym{cf}{CF}{criticality-factor}
\newacronym{mlp}{MLP}{multi-layer-perceptron}
\newacronym{atpg}{ATPG}{Automatic Test Pattern Generation}
\newacronym{ic}{IC}{integrated circuit}
\newacronym{htd}{HTD}{hard-to-detect}
\newacronym{wfp}{WFP}{weighted fault-propagation}

\usepackage{amsthm}
\theoremstyle{plain}

\usepackage{cleveref}
\crefname{enumi}{Step}{Steps}
\crefname{section}{Section}{Sections}
\crefname{subsection}{Section}{Sections}
\crefname{figure}{Figure}{Figures}
\crefname{algocf}{Algorithm}{Algorithms}
\crefname{algorithm}{Algorithm}{Algorithms}
\crefname{algocf}{Algorithm}{Algorithms}
\crefname{table}{Table}{Tables}
\crefname{equation}{Equation}{Equations}
\crefname{eqnarray}{Eq.}{Eq.}
\crefname{appendix}{Section}{Sections}

\tikzstyle{startstop} = [rectangle, rounded corners, 
minimum width=3cm, 
minimum height=1cm,
text centered, 
text width=3.5cm, 
draw=black, 
fill=gray!30]
\tikzstyle{process} = [rectangle, 
minimum width=3cm, 
minimum height=1cm, 
text centered, 
text width=3cm, 
draw=black, 
fill=gray!30]
\tikzstyle{ml_model} = [rectangle, 
minimum width=3cm, 
minimum height=2cm, 
text centered, 
text width=2.5cm, 
draw=black, 
fill=white!100]
\tikzstyle{io} = [trapezium, 
trapezium stretches=true, 
trapezium left angle=70, 
trapezium right angle=110, 
minimum width=3cm, 
minimum height=1cm, 
text centered, 
text width=3cm, 
draw=black, 
fill=gray!30]
\tikzstyle{arrow} = [thick,->,>=stealth]

\newcolumntype{P}[1]{>{\centering\arraybackslash}p{#1}}

\def\BibTeX{{\rm B\kern-.05em{\sc i\kern-.025em b}\kern-.08em
    T\kern-.1667em\lower.7ex\hbox{E}\kern-.125emX}}

\usepackage{fancyhdr}
\renewcommand{\headrulewidth}{0pt}
\makeatletter
\def\@mkbibcitation{\relax}
\renewcommand\footnoterule{
  \kern-3\p@
  \hrule\columnwidth
  \kern2.6\p@}
\makeatother

\usepackage[normalem]{ulem}
\begin{document}

\title{DT-MATCH: \uline{D}efect- and \uline{T}opology-Aware Chiplet \uline{Match}ing for Yield Optimization in Heterogeneous Integration
}
\author{
\IEEEauthorblockN{Xuanyi Tan$^{1}$, Sanmitra Banerjee$^{1}$, Dhruv Thapar$^{1}$, Sreejit Chakravarty$^{2}$ and Krishnendu Chakrabarty$^{1}$}
\IEEEauthorblockA{$^1$School of Electrical, Computer and Energy Engineering, Arizona State University, Tempe, AZ\\
$^2$Ampere Computing, Santa Clara, CA\\
\vspace{-1.5em}
}
}

\maketitle
\thispagestyle{fancy}

\begin{abstract}
Heterogeneous integration (HI) and advanced packaging enable chiplet-based system-on-chip (SoC) designs; however, they also introduce new yield challenges arising from manufacturing defects and complex die-to-die (D2D) interconnect constraints. In particular, lane-level defects, limited spares, and strict structural compatibility requirements impede the efficient assembly of functional systems from defective but repairable chiplets.
We present DT-MATCH, a defect- and topology-aware chiplet matching framework that maximizes the number of functional SoCs under realistic assembly constraints. DT-MATCH provides two complementary approaches. 
The integer linear programming (ILP)-based approach is implemented using a batching strategy to approximate the exact formulation. 
The greedy randomized adaptive search procedure (GRASP)-based approach provides substantial acceleration, but it does not guarantee yield optimality. Both approaches use a preprocessing pipeline to reduce the problem size.
Experimental results on industrial designs demonstrate that, as expected, the ILP-based approach achieves superior solution quality. The GRASP-based approach leads to lower runtimes while still achieving solution quality that is comparable to the ILP-based solution. Together, these methods provide a trade-off between solution quality and efficiency for practical chiplet assembly optimization.

\end{abstract}



\vspace{-0.5em}
\section{Introduction}
\vspace{-0.5em}
Chiplet-based integration has emerged as a practical path for extending system capability when a single monolithic die becomes increasingly constrained by reticle limits, cost, yield, and the need to mix process technologies. Recent industry efforts such as Universal Chiplet Interconnect Express (UCIe) indicate a shift toward an open ecosystem, where reusable chiplets communicate through standardized on-package interconnects, and the package increasingly functions as the system \cite{onufryk2024ucie, huynh2025ucie}. In this setting, the value of a chiplet is determined not only by whether it passes wafer sort, but also by whether a physical instance can be integrated with other available instances to form a complete and functional system. \par

Modern chiplet-based packages make testing and repair a first-order concern. Large numbers of inter-die interconnects are implemented through structures such as micro-bumps and copper pillars, and these links are susceptible to manufacturing defects such as opens and shorts \cite{marinissen2024ieee, chuang2023effective}. The ongoing IEEE P3405 effort is standardizing chiplet interconnect test and repair through dedicated architectural support, including lane numbering, clustering, and redundancy \cite{marinissen2024ieee}. As post-wafer test data becomes richer and more structured, a natural question arises: how should a manufacturer use interconnect defect data before assembly to maximize the number of sellable products? \par

This question is related to yield-oriented matching in three-dimensional (3D) integration \cite{taouil2015yield}. Earlier studies show that selective matching of dies or wafers can improve compound yield in stacked memories and wafer-to-wafer (W2W) bonding \cite{jiang2010yield}. Subsequent work extends these ideas to redundancy sharing, parametric die selection, and assembly-aware planning for heterogeneous bonding \cite{park2012matching, xu2012yield, taouil2015yield}. These solutions established a principle: once pre-bond test data is available, the matching strategy plays a key role in determining final yield. However, these formulations rely on assumptions that do not hold in chiplet-based heterogeneous integration (HI). They fail to capture fine-grained interconnect constraints. As a result, the compatibility among tested chiplets to be assembled is often oversimplified, which does not accurately reflect feasibility in practical chiplet assembly scenarios.

\begin{figure}[t]
    \centering
    \includegraphics[width=0.8\linewidth]{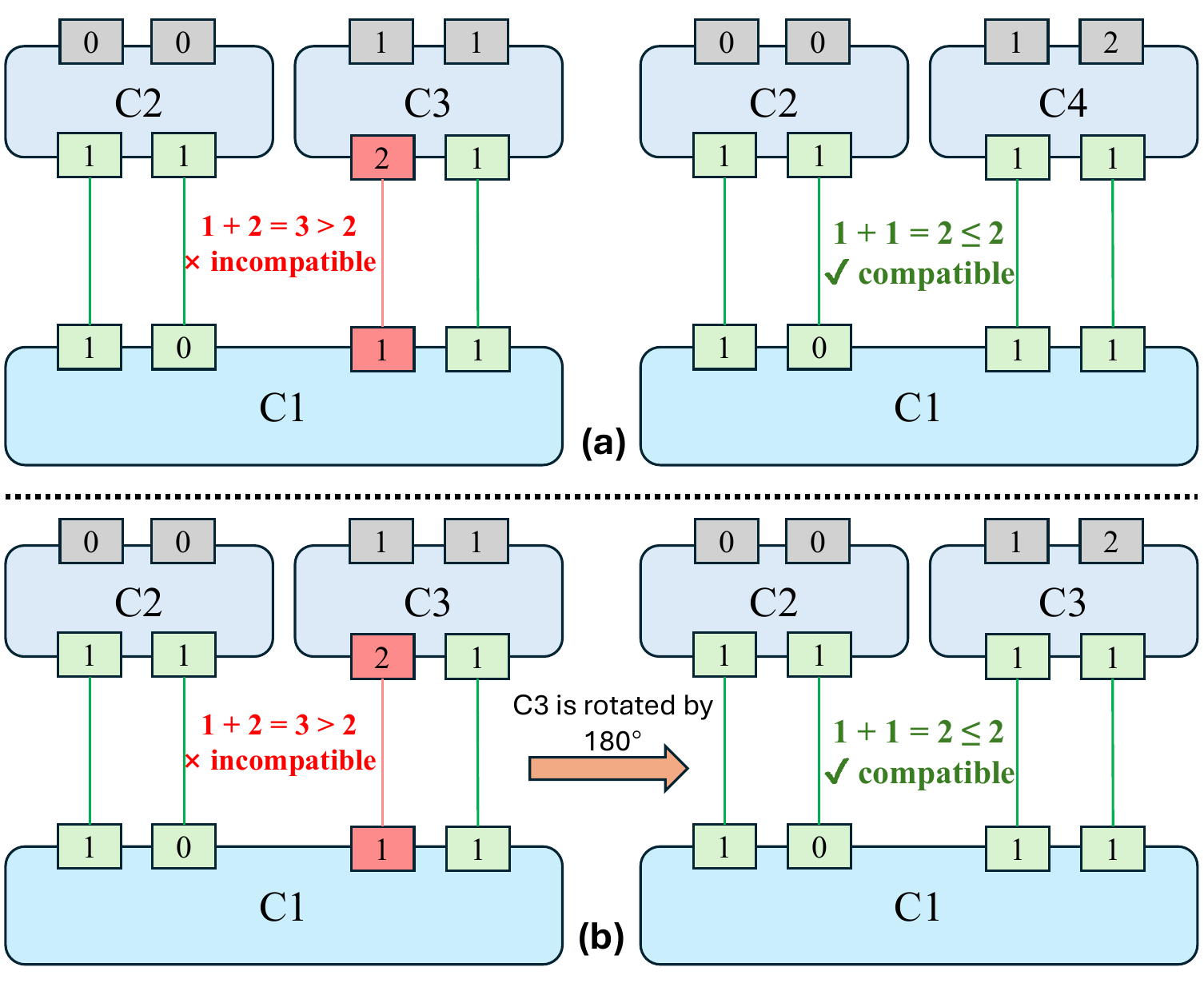}
    \vspace{-0.8em}
    \caption{Motivating examples of defect-aware chiplet assembly. In both scenarios, the default placement (left) causes the combined defect count at the interfaces between C3 and C1 to exceed the shared spare budget of 2. Small squares on each chiplet side denote interfaces, with numbers indicating defect counts (i.e., the number of faulty lanes). Two strategies restore compatibility (right): (a) replacing C3 with another chiplet C4 of the same type but with a more favorable defect distribution; and (b) rotating C3 by 180$^\circ$ so that its less defective side faces C1.}
    \label{fig:example}
    \vspace{-1em}
\end{figure}

In this paper, we present DT-MATCH, a defect- and topology-aware framework for chiplet matching under realistic assembly constraints. We formulate the problem as an assembly optimization over a finite inventory of tested chiplets with defective but repairable interfaces, with the objective of maximizing the number of functionally complete system-on-chip (SoC) instances. Each SoC consists of a set of roles, and each role must be assigned a physical chiplet instance of the required chiplet type, such as GPU, IO, or HBM. Each chiplet instance can be used at most once across all assemblies. The connections among chiplet instances must satisfy a shared spare-budget constraint at each interface position. For a symmetric chiplet whose two orientations are functionally equivalent, a 180° rotation can expose a less defective side and make an otherwise incompatible connection possible when the number of defects at an interface in the original orientation exceeds the allowed spare budget. \cref{fig:example} illustrates a motivating example of chiplet matching. \par

The chiplet matching problem lies at the intersection of test, repair, and optimization. Defect maps and spare-budget information are generated upstream during post wafer level test, while the downstream task is to use this information to optimize the assembly of chiplets into SoCs. Based on this formulation, DT-MATCH includes an exact integer linear programming (ILP) model and a scalable optimization framework that link pre-bond test information to package-level assembly decisions. \par
\par

The contributions of this paper are as follows:
\begin{itemize}
\item We formulate a chiplet matching problem for HI based on post wafer-level test data that captures realistic assembly constraints.
\item We propose a unified lossless preprocessing pipeline to reduce problem size while preserving optimality.
\item We develop two complementary solution approaches. The ILP-based approach achieves high solution quality and is well suited for small- to medium-scale chiplet assembly. The greedy randomized adaptive search procedure (GRASP)-based metaheuristic shows faster runtime while achieving comparable solution quality compared to the ILP-based approach, and is well suited for large-scale instances.
\end{itemize}

The remainder of the paper is organized as follows. \cref{sec:background} provides background and motivation. \cref{sec:problem} describes the preliminaries and problem formulation. \cref{sec:framework} describes the proposed framework. \cref{sec:exp} evaluates the effectiveness of this framework. \cref{sec:conclusion} concludes the paper. \par

\section{Related Prior Work and Research Gap}
\label{sec:background}

\subsection{Yield-Aware Wafer/Die Matching}

Yield enhancement based on post wafer-level test matching has been studied for 3D integration. 
Yield-oriented models for stacked integrated circuits (ICs) are developed in \cite{taouil2015yield}, analyzing the effects of inventory size, matching strategies, and defect assumptions. Prior work shows yield-aware matching can be formulated as a combinatorial optimization problem \cite{dokka2012approximation, duvillie2015complexity}.

Other work studies defect-aware die selection with repair. In 3D-stacked memories, complementary defect patterns enable repairable stacks under redundancy \cite{jiang2010yield, lee2012yield}, and subsequent studies extend these ideas to redundancy allocation and parametric die selection \cite{taouil2012yield, park2012matching, zhang2014novel, xu2012yield, kang2013selection, chen2025yap}. In addition, \cite{zhuang2023multi} formulates multi-product planning with die-to-wafer bonding as an ILP problem. These studies indicate that yield depends on assembly decisions. 
However, prior models assume regular structures (e.g., stacked dies or homogeneous tiers) and do not capture the heterogeneity of chiplets, the presence of multiple connections per chiplet, or compatibility constraints that depend on interface-specific defects under limited spare budgets.

\subsection{Chiplet Interconnect Test, Repair, and Standards}

Test data plays a key role in enabling defect-aware assembly. In 3D stacked ICs, die-centric DfT and IEEE Std 1838 support pre-, mid-, and post-bond testing, treating inter-die interconnects as explicit test targets \cite{marinissen2012dft, 9036129}. These methods emphasize that manufacturing must preserve test access and diagnostic resolution before bonding, a requirement that also applies to chiplet systems with dense micro-bumps and short die-to-die (D2D) links. Recent studies further develop interconnect test methods and architectures, including standardized support such as IEEE P3405 \cite{chuang2023generating, wang2024test, lai2024connectivity, Bhoumik2025SMART}. Related work also models defects and repair under realistic conditions \cite{xie2012yield, bhoumik2025fault}.

Test outputs provide more than binary pass or fail labels. A chiplet may have multiple interfaces, each with different defect counts and remaining spare capacity, so whether it can be used depends on how it is paired with other chiplets. However, these methods mainly focus on detecting and repairing defects, and do not address how tested chiplets should be selected and assembled into complete systems.

\begin{figure*}[t]
    \centering
    \includegraphics[width=0.9\linewidth]{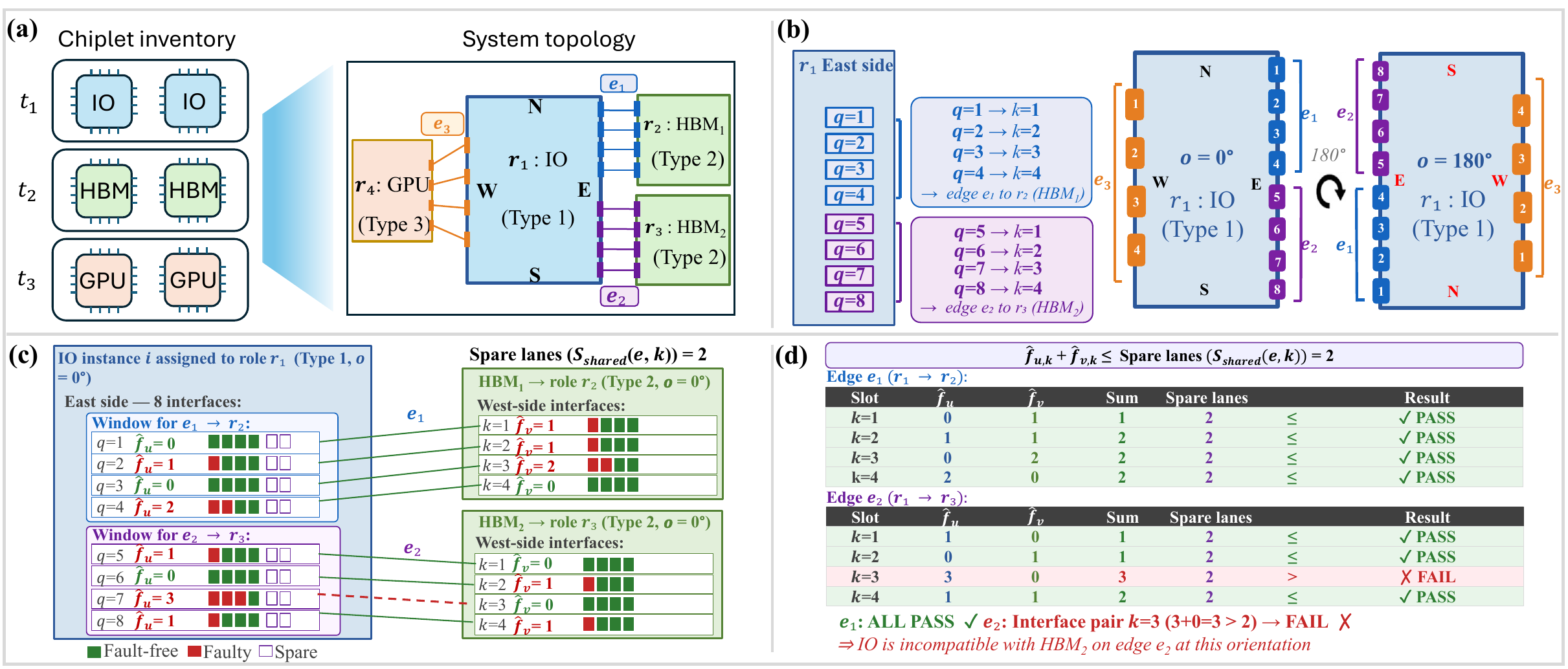}
    \caption{System model illustrated on a simplified SoC example \cite{smith202411}.
    (a)~SoC topology with four roles and topology edges.
    (b)~Disjoint window assignment on the east side of~$r_1$. Effect of $180^{\circ}$ rotation on interface indices.
    (c)~Defect maps and lane-level visualization across edge-local indices for a candidate chiplet assembly.
    (d)~Compatibility verification under Equation~\eqref{eq:pair_compat}.}
    \label{fig:pre}
    \vspace{-1.5em}
\end{figure*}

\subsection{Gap Relative to Prior Work}

This paper addresses a critical gap in SoC assembly at the intersection of test, repair, and optimization. 
Post wafer-level test assembly decisions can improve yield, while test data provides detailed information on D2D link conditions, leading to a constrained assembly optimization problem in HI. 
In this problem, chiplet instances must be selected and placed within a fixed SoC topology, where all required connections must satisfy shared spare-budget constraints, rotations may be applied under design limitations, and each instance can be used at most once. \par

These characteristics are not captured by prior methods, which typically assume regular structures, homogeneous stacks, or wafer-level operations. 
To the best of our knowledge, prior work has not formulated the problem of maximizing the number of assemblable heterogeneous SoCs based on post wafer-level test data under fine-grained defect constraints and limited spare budgets. Our work establishes a direct link between post wafer-level test data and package-level yield optimization.

\section{Preliminaries and Problem Formulation}
\label{sec:problem}
\subsection{Preliminaries}
\label{subsec:prelim}

\subsubsection{SoC Structure and Role Assignment}
We consider a target SoC composed of $m$ chiplets. Since these chiplets serve different functions, we model them as $m$ distinct roles $(r_1, r_2, \ldots, r_m)$. Different roles may correspond to the same chiplet type, for example, multiple roles may require chiplets of the HBM or GPU type. To capture this relation, we define a mapping function $t = type(r)$ that specifies the chiplet type $t$ required by role $r$.
After wafer sort, each chiplet type has an inventory containing all tested instances of that type. The problem is to select instances from these inventories, assign them to roles, and assemble them according to the SoC topology to form a complete SoC.
During assembly, roles must be physically connected. We represent the SoC topology as a set of edges $e$ between roles, where each edge specifies a required connection between the chiplets assigned to two roles. Each chiplet has four physical directions, $d \in \{E, S, W, N\}$, and each edge corresponds to a connection between a specific side of one chiplet and a specific side of another chiplet.

{\flushleft{\underline{Example 1}}}: \cref{fig:pre}(a) illustrates the above model using a simplified SoC with four roles. The design requires two HBM chiplets, one IO chiplet, and one GPU chiplet. For each chiplet type, we assume the inventory contains two tested instances for matching and selection.
The SoC topology specifies the connectivity between roles. The east interfaces of the IO chiplet are partitioned into two groups: one connects to the west interfaces of $HBM_1$, and the other to the west interfaces of $HBM_2$. The west interfaces of the IO chiplet connect to the east interfaces of the GPU chiplet. These connections define the topology and physical interconnections among roles. \hfill $\Box$

\subsubsection{Interface-Level Interconnection Modeling}
\label{sec:im}
To characterize interconnections between chiplets, i.e., edges $e$, we use a finer-grained description of chiplet interfaces. For each direction $d \in \{E, S, W, N\}$, all interfaces along that side are indexed by their spatial order as $(1, 2, \ldots, q)$, where each index corresponds to a physical location.
In practical SoC topologies, an edge represents a mapping between two groups of interfaces on the corresponding sides of two chiplets. For each edge, we model this as selecting a contiguous segment of interface positions on both sides, referred to as a connection window. Within each edge, a local index $k$ denotes the $k$-th interface pair in the window, capturing the relative position within that segment. Interface segments assigned to different edges are non-overlapping, reflecting resource allocation in physical assembly. 
During assembly, chiplets may be placed with different orientations. Without orientation, assigning a role involves selecting a chiplet instance from the inventory of its type. When orientation is considered, the assignment also determines how the instance is placed. We use a variable $o$ to denote orientation, for example, $o = 0^\circ$ indicates no rotation and $o = 180^\circ$ indicates a $180^\circ$ rotation. For each role $r$, the allowed orientations are given by $O_{type(r), r} \subseteq \{0^\circ, 180^\circ\}$, determined by design constraints. Rotation changes both the physical directions and the ordering of interfaces along each side, for example, swapping east and west and reversing the interface sequence. As a result, an interface indexed as $q$ may correspond to a different physical location after rotation, and matching decisions must account for both interface properties and their ordering under the chosen orientation $o$. \par

{\flushleft{\underline{Example 2}}}: \cref{fig:pre}(b) illustrates the relationship among interface indices, edges, and rotation. We consider the east side of the IO chiplet, which has eight interfaces indexed by spatial order as $q = (1, 2, \ldots, 8)$. In the SoC topology, this side participates in two connections: $q = (1, \ldots, 4)$ connect to $HBM_1$ via edge $e_1$, and $q = (5, \ldots, 8)$ connect to $HBM_2$ via edge $e_2$. Thus, each edge corresponds to a distinct subset of interfaces on the same side, forming a connection window.
For each edge, we define a local index $k$ within its interface subset to denote the connection position. For $e_1$, $q = (1, \ldots, 4)$ map to $k = (1, \ldots, 4)$; for $e_2$, $q = (5, \ldots, 8)$ also map to $k = (1, \ldots, 4)$. The index $k$ is defined per edge and is not global.
When the IO chiplet is rotated by $180^\circ$ ($o = 180^\circ$), its east and west sides are swapped and the interface order is reversed. Consequently, indices defined by $q$ map to new positions, which may change the interface subsets used by each connection. Determining connectivity therefore requires considering both interface indices and their ordering under the chosen orientation.\hfill $\Box$

\subsubsection{Defect-Aware Compatibility}
After post wafer-level testing, each chiplet instance is associated with lane-level defect information at every interface, represented as the number of faulty lanes. For an instance $i$ of chiplet type $t$, we denote this by $f_{t,i,d,q}$, the number of faulty lanes at interface position $q$ on side $d$. For each edge $e$ and local index $k$, the D2D protocol provides a shared spare-lane budget $S_{\mathrm{shared}}(e,k)$, which specifies the maximum number of faulty lanes that can be repaired for the corresponding interface pair.

When two instances are assigned to adjacent roles $r_u$ and $r_v$ with orientations $o_u$ and $o_v$, their interfaces are aligned according to edge $e$ and the index mapping, resulting in faulty-lane counts $\hat{f}_{u,k}$ and $\hat{f}_{v,k}$ at each local index $k$. A connection at index $k$ is compatible if and only if
\begin{equation}
\label{eq:pair_compat}
\hat{f}_{u,k} + \hat{f}_{v,k} \le S_{\mathrm{shared}}(e,k).
\end{equation}
This condition must hold for all $k$ on the edge; otherwise, the edge is incompatible and the two chiplets cannot be connected.

{\flushleft{\underline{Example 3}}}: \cref{fig:pre}(c) illustrates compatibility in a concrete example using the two edges on the east side of the IO chiplet that connect to two HBM chiplets. For both edges, the spare-lane budget is 2 for all local indices $k$. For edge $e_1$ between the IO chiplet and $HBM_1$, the sum of faulty lanes on both sides does not exceed 2 at any $k$, so the edge is compatible. For edge $e_2$ between the IO chiplet and $HBM_2$, the sum at the third local index exceeds the budget, making the edge incompatible and preventing the connection.
\cref{fig:pre}(d) presents the same process in tabular form by evaluating each interface position using Equation~\eqref{eq:pair_compat}, providing a direct view of the compatibility checking procedure.
\hfill $\Box$

\subsection{Problem Formulation}
\label{subsec:formulation}
 
\begin{itemize}
\item \textbf{Given:} The target SoC topology defined by a set of roles $r$ and their connectivity; for each chiplet type $t$, the inventory of tested instances and whether $180^\circ$ rotation is supported; for each tested chiplet instance, the faulty lane information $f_{t,i,d,q}$; for each edge $e$ and its $k$-th local connection position, the spare-lane budget $S_{\mathrm{shared}}(e,k)$.
 
\item \textbf{Determine:} An assignment of chiplet instances and feasible rotations to roles such that complete SoCs are assembled.
 
\item \textbf{Subject to:}
  \begin{enumerate}
    \item[\emph{(i)}]  Every role~$r$ in each assembled SoC
          is filled by a chiplet instance of type
          $type(r)$ at a rotation in
          $O_{type(r),r}$.
    \item[\emph{(ii)}] Each chiplet instance can be assigned to at most one SoC.
\item[\emph{(iii)}] For every topology edge and every local interface position, the compatibility condition~(\ref{eq:pair_compat}) is satisfied.
  \end{enumerate}
 
\item \textbf{Objective:} Maximize the total number of complete SoCs.
\end{itemize}

\subsection{Decision Problem and Complexity}
\label{subsec:complexity}

Consider the decision version: given the input and target~$G$, does there exist a feasible assignment of chiplet instances and rotations that assembles at least~$G$ complete SoCs, with all roles filled, each instance used at most once, and Equation~(\ref{eq:pair_compat}) satisfied for every edge and slot?

\begin{theorem}\label{thm:npc}
The decision version of the chiplet matching problem is NP-complete. It remains NP-hard even when each role has a single rotation, each side connects to at most one neighbor, and defect counts and spare budgets are binary.
\end{theorem}

\begin{proof}
Given a candidate solution, we can check in polynomial time that each role is assigned a chiplet of the required type, each chiplet instance is used at most once, and Equation~(\ref{eq:pair_compat}) holds for all connections. Therefore, the problem is in NP.
To prove NP-hardness, we reduce from 3-Dimensional Matching (3DM)~\cite{hartmanis1982computers}. Given three disjoint sets $U,V,W$ of size $n$ and a set of triples $\Gamma \subseteq U \times V \times W$, the problem asks whether there exist $G$ pairwise disjoint triples. We map each element to a leaf chiplet and each triple to a hub chiplet, assigning faulty-lane patterns so that Equation~(\ref{eq:pair_compat}) holds if and only if a hub connects to its three corresponding leaves.

\textbf{Construction.}
We construct a star-shaped SoC with one hub role $r_H$ and three leaf roles $r_U,r_V,r_W$, connected by edges $e_U,e_V,e_W$. Only orientation $0^\circ$ is allowed. Each edge has $n$ interface positions with a shared spare budget of $1$.
For each element $u_i \in U$, we create a leaf chiplet whose faulty-lane vector on edge $e_U$ is $\mathbf{e}_i$; similarly for $V$ and $W$. For each triple $\gamma=(u_i,v_j,w_l)\in \Gamma$, we create a hub chiplet whose three sides have faulty-lane vectors $\mathbf{1}-\mathbf{e}_i$, $\mathbf{1}-\mathbf{e}_j$, and $\mathbf{1}-\mathbf{e}_l$.

\textbf{Correctness.}
On edge $e_U$, the hub corresponding to $\gamma$ is compatible with $u_i$ since $\mathbf{e}_i+(\mathbf{1}-\mathbf{e}_i)=\mathbf{1}$ satisfies Equation~(\ref{eq:pair_compat}). For any $u_{i'} \neq u_i$, the sum exceeds $1$, violating the constraint. The same holds for $e_V$ and $e_W$. Thus each hub can connect only to the three chiplets in its triple, and each feasible SoC corresponds to one triple.
Since each chiplet instance is used at most once, constructing $G$ SoCs is possible if and only if $\Gamma$ contains $G$ disjoint triples. The reduction is polynomial, so the problem is NP-hard. Combined with NP membership, the problem is NP-complete.
\end{proof}

\section{Proposed Optimization Framework}
\label{sec:framework}

\begin{figure}[t]
    \centering
    \includegraphics[width=0.9\linewidth]{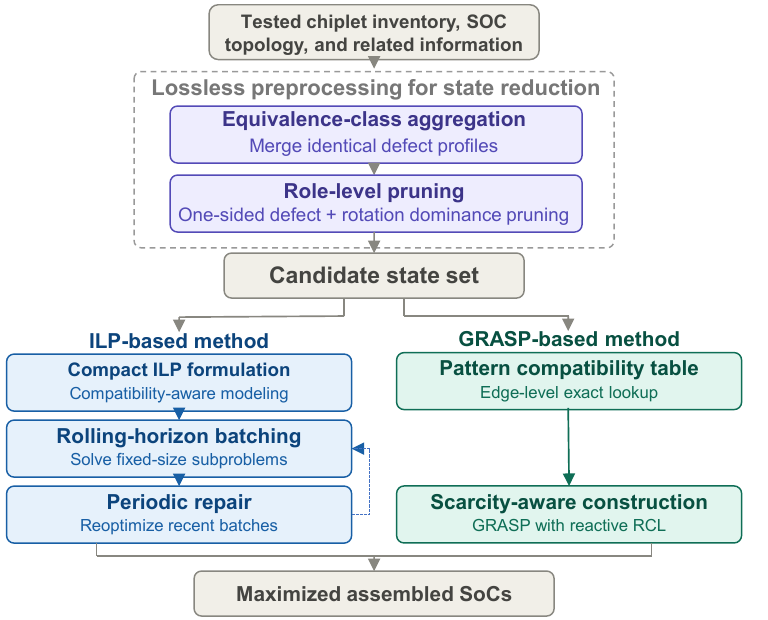}
    \caption{Overview of the DT-MATCH framework.}
    \label{fig:flow}
    \vspace{-1.5em}
\end{figure}

The NP-completeness result of Theorem~\ref{thm:npc} motivates two complementary solution approaches. Fig.~\ref{fig:flow} provides an overview of the framework. Both approaches share a lossless preprocessing pipeline (Section~\ref{sec:scalable}) that reduces the problem size. The approaches then diverge: one adopts an ILP-based formulation with rolling-horizon batching, while the other employs a GRASP-based heuristic for a scalable solution.

\subsection{Exact Preprocessing}
\label{sec:scalable}

\subsubsection{Equivalence-Class Aggregation}
\label{sec:eca}

We compress the inventory of each chiplet type by aggregating equivalent instances. Two instances of the same type are considered equivalent if they have identical faulty-lane values at all interface positions relevant to feasible assembly.

The aggregation has two steps. First, we retain only the interface positions used in the SoC topology. For each retained position $(d,q)$ of type $t$, let $B^{\max}_{t,d,q}$ be the largest spare budget that can occur at that position. For each instance $i$ and each retained position $(d,q)$, we clip the faulty-lane value as
$\tilde{f}_{t,i,d,q}
= \min\!\bigl\{f_{t,i,d,q},\;B^{\max}_{t,d,q}+1\bigr\}$,
since larger values cannot be repaired and are therefore indistinguishable.
Second, instances with identical clipped values at all retained positions are grouped into equivalence classes $c$. Each class represents a unique faulty-lane pattern and contains all instances with that pattern. For each class $c$ of type $t$, we record the number of instances it contains, denoted by $m_{t,c}$.
Before aggregation, each role is assigned a chiplet instance and an orientation $o$, forming a state $(i,o)$. After aggregation, instances within the same class $c$ are interchangeable, so the assignment can be represented as $(c,o)$, referred to as a state. This replaces instance-level states $(i,o)$ with class-level states $(c,o)$, reducing the number of states and thus the size of the search space.

\begin{figure}[t]
    \centering
    \includegraphics[width=0.95\linewidth]{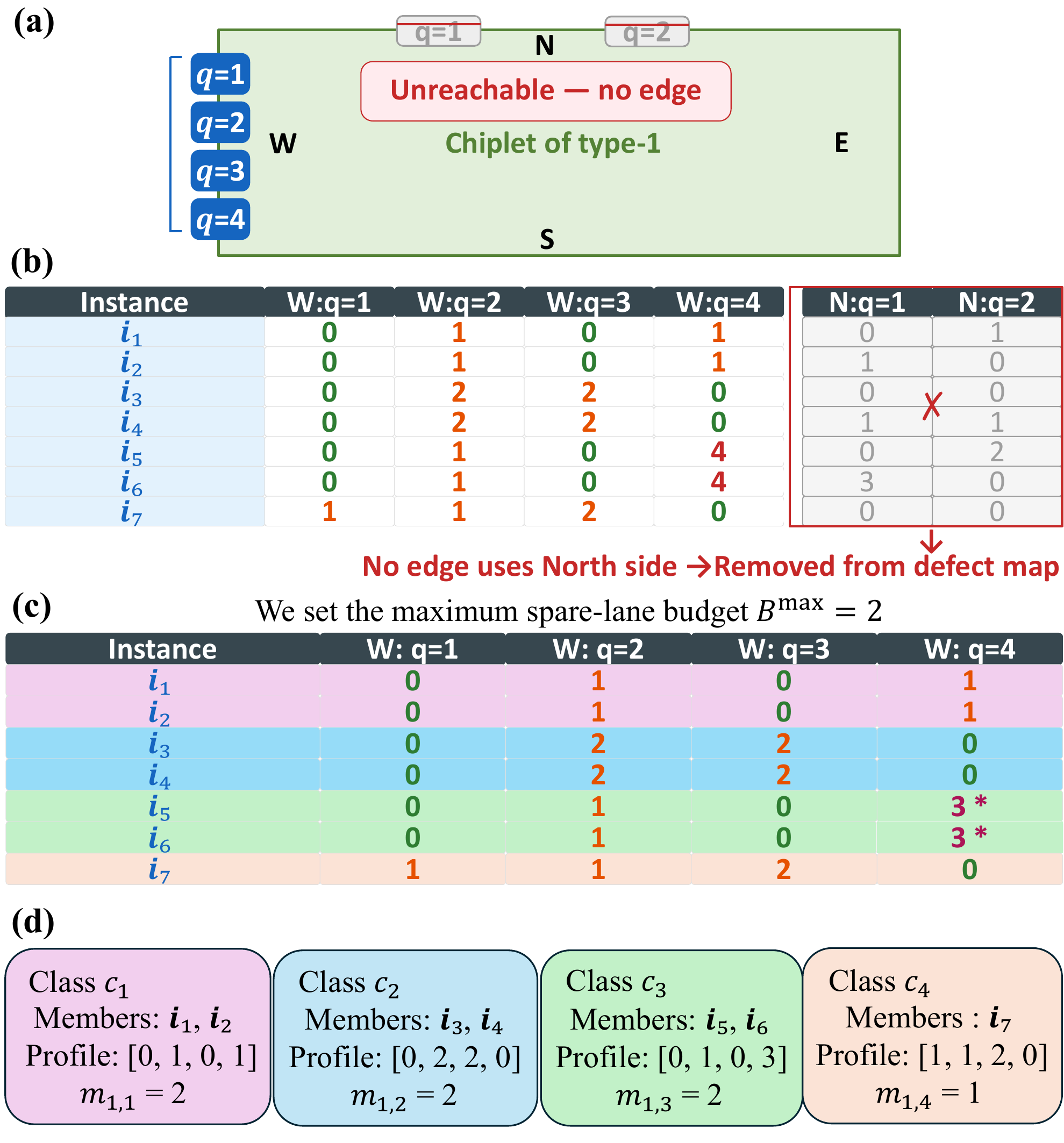}
    \caption{Equivalence-class aggregation illustrated on a type-1 chiplet. (a) Interface layout with reachable and unreachable sides. (b) Raw defect map; North-side columns are removed. (c) Clipped defect profiles with sentinel values marked by *. (d) Resulting four equivalence classes with their instance counts.}
    \label{fig:equal}
    \vspace{-1.5em}
\end{figure}

{\flushleft{\underline{Example 4}}}:
\cref{fig:equal} illustrates the preprocessing procedure on a specific chiplet type. The north-side interfaces do not participate in any connections in the SoC topology, while the west-side interfaces do, so the north side is discarded and only the west side is retained, as shown in \cref{fig:equal}(b).
Based on this reduction, the faulty-lane values on the west-side interfaces are simplified. Suppose the spare budget at each position is 2. When the faulty-lane count exceeds this budget, its exact value is irrelevant and is capped at the budget plus one. For example, for the 5th and 6th instances, the faulty-lane count at the 4th west-side interface is 4, which is capped at 3.
Finally, instances with identical simplified patterns are grouped, partitioning the inventory of this chiplet type into four equivalence classes.
\hfill $\Box$

\subsubsection{Role-Level Pruning}

Before equivalence-class aggregation, the faulty-lane information of each instance is described by $f_{t,i,d,q}$. After aggregation, instances are grouped into classes, so individual instances no longer need to be tracked. Instead, we define a class-level quantity $\alpha_{r,e,k,c,o}$ to represent the number of faulty lanes from class $c$ at the $k$-th interface pair on edge $e$ under orientation $o$. This value is obtained from a representative instance of class $c$ after applying orientation $o$. Since all instances in $c$ share identical clipped faulty-lane values at relevant interface positions, this value applies to the entire class.

In \cref{sec:eca}, we introduce class--orientation states $(c,o)$. However, not all states are useful for a given role, as some can never participate in a feasible assembly. We therefore apply two role-level pruning techniques.
The first, one-sided pruning, removes $(c,o)$ if it already violates the spare-budget constraint, i.e., if $\alpha_{r,e,k,c,o} > S_{\mathrm{shared}}(e,k)$ for some $(e,k)$. In this case, no pairing can make the connection compatible. The second, rotation-dominance pruning, removes $(c,o_2)$ if there exists $(c,o_1)$ such that $\alpha_{r,e,k,c,o_1} \le \alpha_{r,e,k,c,o_2}$ for all $(e,k)$, with strict inequality for at least one position. In this case, $o_2$ is never better than $o_1$. The remaining states are used in subsequent SoC assembly.

\begin{figure}[t]
    \centering
    \includegraphics[width=0.95\linewidth]{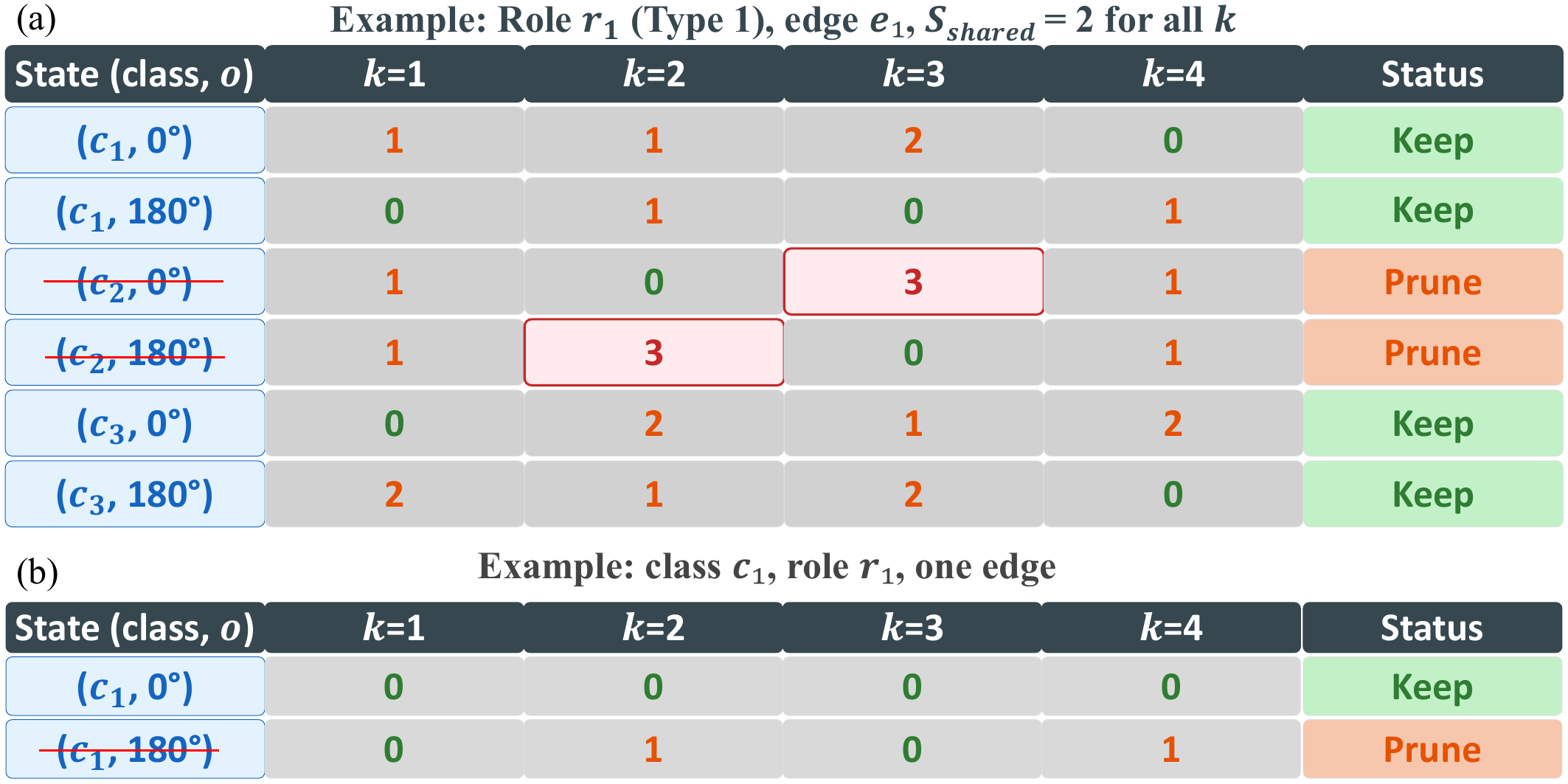}
    \caption{Role-level pruning illustrated on two examples. (a) One-sided defect pruning: states whose defect contribution exceeds the shared spare budget at any slot are pruned. (b) Rotation-dominance pruning: an orientation uniformly dominated by another orientation of the same class is pruned.}
    \label{fig:prune}
    \vspace{-1.5em}
\end{figure}

{\flushleft{\underline{Example 5}}}: Fig.~\ref{fig:prune} illustrates the pruning process. 
In Fig.~\ref{fig:prune}(a), $(c_2,0^{\circ})$ is removed because the number of faulty lanes at $k=3$ exceeds the spare budget. 
The same applies to $(c_2,180^{\circ})$. 
In Fig.~\ref{fig:prune}(b), the orientation $o=180^{\circ}$ has faulty-lane values $[0,1,0,1]$, while $o=0^{\circ}$ has values $[0,0,0,0]$. 
Since $o=180^{\circ}$ is dominated by $o=0^{\circ}$, it is removed.
\hfill $\Box$

Equivalence-class aggregation and role-level pruning together provide a lossless reduction in problem size, reducing the number of variables and constraints in the ILP-based formulation and the search space explored by the GRASP-based method.

\subsection{ILP-Based Method}
\label{sec:ilp}


For each chiplet type, let $N_t$ be the number of available instances and $n_t$ the number of roles that require it. The maximum number of SoCs that can be assembled is bounded by $S_{\max}=\min_t \lfloor N_t/n_t \rfloor$, determined by the most constrained type.
We index assemblies by $s\in\{1,\ldots,S_{\max}\}$. A binary variable $Y_s$ indicates whether assembly $s$ is realized, and $X_{s,r,c,o}$ indicates whether state $(c,o)$ is assigned to role $r$ in assembly $s$. The objective is to maximize the number of assembled SoCs, $\max \sum_s Y_s$.
 
The ILP formulation enforces four groups of constraints:
 
\begin{itemize}

 

\item \textit{Role completeness.} Each role in an assembled SoC must select exactly one surviving state:
$\sum_{(c,o)} X_{s,r,c,o}=Y_s$. When $Y_s=1$, exactly one class-orientation state is selected for each role; when $Y_s=0$, all variables $X_{s,r,c,o}$ are forced to $0$, and the assembly is not formed.

\item \textit{Class usage limit.} Each equivalence class can only be used as many times as available in the inventory:
$\sum_s \sum_{\substack{r:\mathrm{type}(r)=t}} \sum_{o} X_{s,r,c,o} \le m_{t,c}$. 
This constraint limits how many times class $c$ can be assigned across all assemblies. 
Once the available instances in class $c$ are exhausted, no assignments of this class are allowed.

\item \textit{Interface-level defect compatibility.} For every connection, consider an edge $e$ between two roles $r_u$ and $r_v$. 
The total number of faulty lanes from both endpoints must be within the available spare budget:
\vspace{-0.5em}
\begin{multline}
\sum_{(c,o)} \alpha_{r_u,e,k,c,o}\cdot X_{s,r_u,c,o} \\
+\sum_{(c',o')} \alpha_{r_v,e,k,c',o'}\cdot X_{s,r_v,c',o'}
\le S_{\mathrm{shared}}(e,k)\cdot Y_s.
\end{multline}
When $Y_s=1$, the combined faulty-lane values at each connection position must not exceed the spare budget; when $Y_s=0$, the constraint is inactive because no assignments are made.

\item \textit{Symmetry breaking.} Assemblies are activated in order: $Y_s\ge Y_{s+1}$. This enforces that assembly $s+1$ can be used only if assembly $s$ is used, avoiding equivalent solutions.
\end{itemize}
This formulation is exact but becomes expensive for large inventories. To improve scalability, we use a rolling-horizon batching strategy. Instead of solving for all $S_{\max}$ assemblies at once, each iteration solves a smaller ILP over the current remaining inventory. After each iteration, selected assignments are fixed, class counts are updated, and states that become infeasible under the updated inventory are removed before the next batch.
Because decisions are made sequentially, early batches may consume instances that are more valuable later, leading to suboptimal yield. To mitigate this, we periodically revisit recent batches and reoptimize them: assignments in these batches are released and recomputed jointly under the same ILP constraints, with priority given to chiplet types with the fewest remaining instances. The updated solution is accepted only if it increases the total number of assembled SoCs, allowing the solver to correct earlier decisions while keeping the problem size manageable.

\subsection{GRASP-Based Method}


The GRASP-based method follows the standard GRASP framework~\cite{feo1995greedy}, which consists of multiple independent iterations. 
Each iteration constructs one solution, and the best solution across all iterations is retained.
At a high level, each iteration starts from the full chiplet inventory and repeatedly builds complete SoCs. 
Each SoC is constructed step by step by assigning chiplets to roles, while ensuring that all connections satisfy the compatibility condition in Equation~(\ref{eq:pair_compat}). 
Once no more complete SoCs can be formed, the iteration produces a set of assembled SoCs as its solution.

\subsubsection{Preprocessing and Compatibility}
\label{sec:pac}
For each edge, we precompute the compatibility between class--orientation states at its two endpoints and store the results in a compatibility table. A pair is marked compatible if their faulty-lane values satisfy the spare-budget constraint at every interface position. This table is used during construction, where a complete SoC is built step by step by assigning chiplets to roles, to enable fast validity checking.

\subsubsection{Greedy Randomized Construction}

In each GRASP iteration, SoCs are constructed sequentially from the residual inventory, each built incrementally from a seed role along the topology. The seed is chosen to prioritize the most constrained resource: for each chiplet type, we compute the ratio between its remaining instances and the number of roles that require it, and select the role with the smallest ratio.

The SoC is then expanded by assigning adjacent roles. For each role, all $(c,o)$ choices are evaluated, and a choice is feasible if the class $c$ has remaining instances, it is compatible with all assigned neighbors according to the precomputed table in \cref{sec:pac}, and it does not eliminate all feasible $(c,o)$ options for any adjacent unassigned role, avoiding dead ends.

When multiple feasible $(c,o)$ choices exist, a restricted candidate list (RCL) is formed. Candidates are ranked by a scarcity score, where smaller values indicate that the corresponding class has fewer remaining instances, and one $(c,o)$ is selected uniformly at random. The RCL size is adjusted using reactive GRASP~\cite{prais2000reactive}, using smaller lists for stronger guidance and larger ones for more exploration.

{\flushleft{\underline{Example 6}}}: Consider a simple SoC with three roles $r_1$, $r_2$, and $r_3$. Suppose the chiplet type required by $r_1$ is the most constrained, while the types for $r_2$ and $r_3$ have larger inventories. The role $r_1$ is therefore selected as the seed. Assume two feasible states exist for $r_1$, $(c_a,0^{\circ})$ and $(c_b,180^{\circ})$, both satisfying the feasibility conditions under the current partial assignment. If class $c_a$ has fewer remaining instances than $c_b$, then $(c_a,0^{\circ})$ receives a better scarcity score. Both may enter the RCL, from which one is selected. \hfill $\Box$

The algorithm then proceeds to an adjacent role with the fewest feasible $(c,o)$ options and continues until all roles are assigned, yielding one complete SoC.
Once all the roles have been assigned, a single physical instance is selected from each chosen class, and the corresponding inventory counts are decremented accordingly. This step completes the construction of one SoC. The method then attempts to construct an additional SoC using the updated residual inventory. This iterative process continues until no further complete SoC can be assembled, ultimately yielding a set of SoCs for the current iteration.



The overall procedure is summarized in Algorithm~\ref{alg:grasp}. The GRASP-based method does not solve a global optimization model, but it scales better on large inventories.

\begin{algorithm}[t]
\caption{GRASP-Based SoC Assembly}
\label{alg:grasp}
\small
\setlength{\baselineskip}{9pt}
\begin{algorithmic}[1]
\Require
\Statex \makebox[3.5em][l]{$A_r^{\mathrm{class}}$} // candidate class-orientation pairs for each role
\Statex \makebox[3.5em][l]{$m_{t,c}$}              // multiplicity of class $c$ of type $t$
\Statex \makebox[3.5em][l]{$N_{\mathrm{iter}}$}     // number of GRASP iterations
\Ensure \makebox[3.5em][l]{$X_{\mathrm{best}}$}     // best solution
\State $X_{\mathrm{best}} \gets \emptyset$
\For{$\ell = 1$ to $N_{\mathrm{iter}}$}
    \State initialize residual inventory
    \State $X \gets \emptyset$
    \While{a complete SoC can be constructed}
        \State initialize empty SoC $S$, select seed role
        \While{$S$ is incomplete}
            \State select next role (most constrained)
            \State compute feasible candidates, build RCL and sample one candidate, assign and update inventory
        \EndWhile
        \State add $S$ to $X$
    \EndWhile
    \If{$|X| > |X_{\mathrm{best}}|$}
        \State $X_{\mathrm{best}} \gets X$
    \EndIf
\EndFor
\State \Return $X_{\mathrm{best}}$
\vspace{-3pt}
\end{algorithmic}

\end{algorithm}

\section{Experimental Evaluation}
\label{sec:exp}
\subsection{Experimental Setup}
All experiments for DT-MATCH are conducted on a workstation with AMD EPYC 7313 processors. The ILP model is implemented in Python and solved using Gurobi via its Python API with default parameters \cite{gurobi}. Parameter settings for both the ILP-based and GRASP-based methods are chosen based on a sweep to balance solution quality and runtime. For the ILP-based method, the batch size is 150. The repair procedure is applied every two batches and reoptimizes the last four batches. For the GRASP-based method, the number of iterations is 5.

\subsection{Evaluation Metrics}
To evaluate the proposed methods, we compare yield and runtime under different problem sizes. For a fixed SoC topology, the problem size is characterized by $S_{\max}$, the upper bound on the number of complete SoCs that can be assembled from the available inventory. Since different inventories can share the same $S_{\max}$, we generate multiple inventory instances for each $S_{\max}$ to capture variation in composition while keeping the problem size fixed. Here, an inventory instance refers to a specific configuration of chiplet counts across chiplet types, rather than individual chiplet instances.
Specifically, we introduce a $20\%$ variation in the number of chiplets of each type. For each inventory instance, one type is randomly selected as the bottleneck that determines $S_{\max}$, while the remaining types are assigned with small random variations, ensuring none becomes more restrictive. As a result, all inventory instances share the same $S_{\max}$ but differ in composition.

For each $S_{\max}$, we generate $N$ inventory instances, where larger $N$ increases diversity and smaller $N$ reduces runtime overhead. For each method, we report yield and runtime across the $N$ instances. Yield, defined as the ratio of assembled SoCs to $S_{\max}$, is summarized by its mean, median, minimum, and maximum, denoted by $Y_{\text{mean}}$, $Y_{\text{med}}$, $Y_{\min}$, and $Y_{\max}$. Runtime is reported similarly as $T_{\text{mean}}$, $T_{\text{med}}$, $T_{\min}$, and $T_{\max}$. This setup captures both solution quality and efficiency, as well as variability across inventory instances.


\subsection{Benchmarks}
\begin{table}[t]
\centering
\caption{Published benchmark statistics.}
\vspace{-0.5em}
\label{tab:bench}
\footnotesize
\renewcommand{\arraystretch}{0.1}
\setlength{\tabcolsep}{2pt}
\begin{tabular}{|c|c|c|c|c|}
\hline
\textbf{ID} & \textbf{Benchmarks} & \textbf{Description} & \textbf{Chiplet composition} & \textbf{\# Chiplets} \\
\hline
B & \makecell[c]{AMD Instinct\\MI300X ~\cite{smith202411}}
  & \makecell[c]{An all-GPU\\data center\\accelerator}
  & \makecell[c]{$8\,\mathrm{XCD} + 4\,\mathrm{IOD}$\\$+ 8\,\mathrm{HBM3}$}
  & 20 \\
\hline
A & \makecell[c]{AMD Instinct\\MI300A ~\cite{amd_mi300_ieee_spectrum}}
  & \makecell[c]{AMD's first\\data center\\APU}
  & \makecell[c]{$6\,\mathrm{XCD} + 3\,\mathrm{CCD}$\\
               $+ 4\,\mathrm{IOD} + 8\,\mathrm{HBM3}$}
  & 21 \\
\hline
\end{tabular}
\vspace{-0.5em}
\end{table}
               
\begin{table}[t]
\centering
\caption{Proprietary industrial benchmark statistics.}
\vspace{-0.5em}
\label{tab:bench_2}
\footnotesize
\renewcommand{\arraystretch}{0.5}
\setlength{\tabcolsep}{0.1pt}
\begin{tabular}{|c|c|c|c|c|c|c|c|}
\hline
\multirow{2}{*}{\textbf{ID}} 
& \multirow{2}{*}{\textbf{Chiplet composition}} 
& \multicolumn{3}{c|}{\textbf{Unit die cost}} 
& \multirow{2}{*}{\textbf{SoC cost}} 
& \multirow{2}{*}{\textbf{Package}} 
& \multirow{2}{*}{\textbf{Total}} \\
\cline{3-5}
 &  & \textbf{C1} & \textbf{C2} & \textbf{C3} 
 &  &  &  \\
\hline
D1 & \makecell[c]{$\mathrm{C1}_{d1} + 2\,\mathrm{C2}_{d1}$\\
                 $+\,2\,\mathrm{C3}_{d1}$}
   & \$1,000 & \$250 & \$250 
   & \$2,000 & \$500 & \$2,500 \\
\hline
D2 & \makecell[c]{$\mathrm{C1}_{d2}$\\
                 $+\,8\,\mathrm{C2}_{d2}$}
   & \$1,000 & \$2,000 & - 
   & \$17,000 & \$500 & \$17,500 \\
\hline
\end{tabular}
\vspace{-2em}
\end{table}

\begin{table*}[t]
    \centering
    \caption{Comparison of five methods across benchmarks for DT-MATCH. Yield is reported as \( (Y_{\min}, Y_{\text{med}}, Y_{\max}) \), and runtime is reported as \( (T_{\min}, T_{\text{med}}, T_{\max}) \).}
    \renewcommand{\arraystretch}{0.95}
    \setlength{\tabcolsep}{1.5pt}
    \scriptsize
    \begin{tabular}{|c|c|c|c|c|c|c|c|c|c|}
        \hline
        \multirow{2}{*}{\textbf{$S_{\max}$}} &
        \multirow{2}{*}{\textbf{Method}} &
        \multicolumn{2}{c|}{\textbf{Benchmark A}} &
        \multicolumn{2}{c|}{\textbf{Benchmark B}} &
        \multicolumn{2}{c|}{\textbf{Benchmark D1}} &
        \multicolumn{2}{c|}{\textbf{Benchmark D2}} \\
        \cline{3-10}
        & &
        \textbf{Yield (\%)} & \textbf{Runtime (min)} &
        \textbf{Yield (\%)} & \textbf{Runtime (min)} &
        \textbf{Yield (\%)} & \textbf{Runtime (min)} &
        \textbf{Yield (\%)} & \textbf{Runtime (min)} \\
        \hline

        \multirow{5}{*}{200}
        & \textbf{Batched ILP} & (68.5, 75.0, 79.5) & (2.6, 4.5, 17.6) & (71.0, 76.0, 77.0) & (3.2, 3.8, 8.1) & (94.5, 96.5, 97.5) & (0.1, 0.1, 0.1) & (95.5, 97.5, 100.0) & (0.3, 0.3, 0.3) \\
        & \textbf{GRASP}       & (66.5, 71.5, 75.5) & (2.3, 2.4, 2.5) & (69.0, 71.5, 75.5) & (1.7, 2.0, 2.2) & (94.5, 95.0, 97.0) & (0.1, 0.1, 0.1) & (95.5, 97.5, 99.5) & (0.3, 0.3, 0.3) \\
        & Monolithic ILP       & (68.5, 75.0, 80.5) & (6.6, 6.8, 7.3) & (71.0, 76.0, 78.0) & (6.4, 7.1, 7.4) & (94.5, 96.5, 97.5) & (0.1, 0.2, 0.2) & (95.5, 97.5, 100.0) & (0.1, 0.3, 0.4) \\
        & Greedy               & (0.0, 0.0, 0.5) & (0.0, 0.0, 0.0) & (0.0, 0.0, 0.0) & (0.0, 0.0, 0.0) & (43.5, 88.5, 91.0) & (0.0, 0.0, 0.0) & (67.5, 86.5, 88.5) & (0.0, 0.0, 0.0) \\
        & Random               & (0.0, 0.0, 0.0) & (0.0, 0.0, 0.0) & (0.0, 0.0, 0.0) & (0.0, 0.0, 0.0) & (68.5, 74.5, 79.5) & (0.0, 0.0, 0.0) & (55.5, 58.0, 66.0) & (0.0, 0.0, 0.0) \\
        \hline

        \multirow{5}{*}{400}
        & \textbf{Batched ILP} & (71.8, 74.3, 75.8) & (9.9, 10.5, 13.3) & (71.3, 73.8, 76.3) & (8.0, 9.3, 22.3) & (95.3, 97.8, 100.0) & (0.5, 0.5, 0.9) & (96.5, 97.5, 100.0) & (0.8, 0.8, 0.8) \\
        & \textbf{GRASP}       & (69.0, 70.5, 73.8) & (4.4, 5.3, 5.7) & (69.5, 72.3, 73.3) & (4.3, 4.6, 5.1) & (95.0, 97.8, 100.0) & (0.2, 0.2, 0.2) & (96.5, 97.5, 100.0) & (0.6, 0.6, 0.7) \\
        & Monolithic ILP       & (71.8, 74.3, 76.3) & (39.0, 43.6, 46.8) & (72.0, 74.0, 77.8) & (77.4, 82.6, 118.8) & (95.3, 97.8, 100.0) & (0.2, 0.7, 0.7) & (96.5, 97.5, 100.0) & (0.4, 0.8, 1.4) \\
        & Greedy               & (0.0, 0.0, 0.5) & (0.0, 0.0, 0.0) & (0.0, 0.0, 0.3) & (0.0, 0.0, 0.0) & (81.8, 86.0, 94.5) & (0.0, 0.0, 0.0) & (62.5, 77.3, 90.5) & (0.0, 0.0, 0.0) \\
        & Random               & (0.0, 0.0, 0.0) & (0.0, 0.0, 0.0) & (0.0, 0.0, 0.0) & (0.0, 0.0, 0.0) & (72.5, 76.0, 77.8) & (0.0, 0.0, 0.0) & (54.3, 59.0, 64.0) & (0.0, 0.0, 0.0) \\
        \hline

        \multirow{5}{*}{600}
        & \textbf{Batched ILP} & (70.2, 75.0, 77.8) & (23.9, 25.8, 41.1) & (69.5, 72.3, 74.7) & (28.9, 34.4, 43.2) & (96.2, 96.7, 100.0) & (0.7, 0.7, 0.7) & (96.8, 97.3, 100.0) & (0.9, 0.9, 1.0) \\
        & \textbf{GRASP}       & (68.5, 71.0, 75.7) & (7.9, 9.1, 9.7) & (68.3, 71.0, 72.7) & (6.7, 6.9, 7.7) & (95.3, 96.7, 98.7) & (0.3, 0.3, 0.3) & (96.7, 97.3, 99.5) & (0.9, 1.0, 1.0) \\
        & Monolithic ILP       & (70.7, 75.3, 78.3) & (95.7, 118.1, 130.8) & (70.7, 73.8, 75.3) & (494.1, 614.2, 1086.1) & (96.2, 96.8, 100.0) & (0.5, 2.0, 2.6) & (96.8, 97.3, 100.0) & (0.8, 2.1, 2.4) \\
        & Greedy               & (0.0, 0.0, 0.0) & (0.0, 0.0, 0.0) & (0.0, 0.0, 0.5) & (0.0, 0.0, 0.0) & (36.3, 86.8, 93.0) & (0.0, 0.0, 0.0) & (80.0, 85.3, 88.3) & (0.1, 0.1, 0.1) \\
        & Random               & (0.0, 0.0, 0.0) & (0.0, 0.0, 0.0) & (0.0, 0.0, 0.0) & (0.0, 0.0, 0.0) & (73.3, 74.5, 76.0) & (0.0, 0.0, 0.0) & (55.7, 58.3, 61.3) & (0.0, 0.0, 0.0) \\
        \hline

        \multirow{5}{*}{800}
        & \textbf{Batched ILP} & (71.3, 72.1, 81.3) & (36.0, 44.2, 45.5) & (70.5, 71.6, 80.9) & (28.3, 40.5, 42.9) & (96.3, 97.8, 100.0) & (1.1, 1.1, 1.1) & (97.4, 97.5, 100.0) & (1.3, 1.4, 1.5) \\
        & \textbf{GRASP}       & (69.5, 70.5, 75.0) & (10.8, 11.5, 14.7) & (69.4, 70.0, 79.6) & (8.0, 10.0, 10.4) & (96.0, 97.0, 99.5) & (0.3, 0.3, 0.4) & (97.3, 97.5, 99.8) & (1.2, 1.3, 1.4) \\
        & Monolithic ILP       & (71.1, 72.5, 83.0) & (205.9, 248.1, 395.5) & timeout & timeout & (96.3, 97.8, 100.0) & (1.0, 2.2, 2.4) & (97.4, 97.5, 100.0) & (2.5, 4.1, 4.9) \\
        & Greedy               & (0.0, 0.0, 0.1) & (0.0, 0.0, 0.0) & (0.0, 0.0, 0.0) & (0.0, 0.0, 0.0) & (59.5, 89.3, 91.6) & (0.0, 0.0, 0.0) & (12.5, 75.5, 78.9) & (0.0, 0.1, 0.1) \\
        & Random               & (0.0, 0.0, 0.0) & (0.0, 0.0, 0.0) & (0.0, 0.0, 0.0) & (0.0, 0.0, 0.0) & (74.6, 75.9, 77.4) & (0.0, 0.0, 0.0) & (56.8, 57.8, 58.9) & (0.0, 0.0, 0.0) \\
        \hline

        \multirow{5}{*}{1000}
        & \textbf{Batched ILP} & (71.0, 73.9, 79.8) & (49.8, 60.3, 66.2) & (70.0, 71.3, 78.8) & (39.1, 49.9, 67.1) & (96.2, 96.3, 100.0) & (3.7, 4.9, 5.2) & (96.4, 97.1, 100.0) & (5.4, 5.6, 6.9) \\
        & \textbf{GRASP}       & (69.3, 70.3, 76.4) & (13.6, 15.1, 15.6) & (68.9, 70.3, 75.6) & (11.6, 12.0, 12.8) & (95.3, 96.3, 100.0) & (0.4, 0.4, 0.5) & (96.5, 97.1, 99.9) & (1.7, 1.7, 1.7) \\
        & Monolithic ILP       & (71.5, 74.3, 80.2) & (500.3, 624.9, 928.3) & timeout & timeout & (96.3, 96.3, 100.0) & (1.1, 3.2, 3.4) & (96.5, 97.1, 100.0) & (1.6, 7.5, 8.8) \\
        & Greedy               & (0.0, 0.0, 0.0) & (0.0, 0.0, 0.0) & (0.0, 0.0, 0.1) & (0.0, 0.0, 0.0) & (78.0, 78.5, 92.9) & (0.0, 0.0, 0.0) & (57.4, 70.5, 89.8) & (0.1, 0.1, 0.2) \\
        & Random               & (0.0, 0.0, 0.0) & (0.0, 0.0, 0.0) & (0.0, 0.0, 0.0) & (0.0, 0.0, 0.0) & (72.9, 75.7, 77.0) & (0.0, 0.0, 0.0) & (57.6, 58.9, 59.2) & (0.0, 0.0, 0.0) \\
        \hline
    \end{tabular}
    \label{tab:benchmark_results}
    \vspace{-1.5em}
\end{table*}

We evaluate our method on four benchmarks. Two of these, shown in \cref{tab:bench}, are constructed from published industry papers, and the other two, shown in \cref{tab:bench_2}, are anonymized real designs from industry.

Each design consists of multiple chiplets as specified in the chiplet composition column. For example, D1 includes five chiplets (one $C_1$, two $C_2$, and two $C_3$), while D2 includes nine chiplets (one $C_1$ and eight $C_2$). In \cref{tab:bench_2}, the unit die cost columns ($C_1$, $C_2$, $C_3$) report the cost per chiplet type. The SoC cost is the sum over all chiplets in a design, and the total cost further includes the package cost. These values support the yield-aware cost analysis in \cref{sec:cost}, which quantifies the financial impact of yield loss.

For all benchmarks, faulty lanes at each interface of every chiplet instance are generated using a Poisson distribution with parameter $\lambda=1$, applied uniformly across chiplet types, instances, directions, and interface positions. In \cref{sec:main}, for each $S_{\max}$, we generate $N=20$ inventory instances. Each inventory instance uses a different random seed in the Poisson sampling, introducing variability across inventory instances.

\subsection{Yield and Runtime Results} 
\label{sec:main}


We compare the ILP-based approach and the GRASP-based approach with three other methods on four benchmarks, including monolithic ILP, naive random matching, and naive greedy. The three methods are described as follows.
\begin{itemize}
\item Monolithic ILP: this method removes the rolling-horizon batching from our ILP-based approach and solves the entire problem in a single formulation, which serves as a proxy for optimal solutions.
\item Random matching: this method assigns chiplet instances to roles uniformly at random without considering any constraints.
\item Greedy: this method traverses roles in a fixed order and, for each role, selects the first $(c,o)$ in the instance sequence that is compatible with all already assigned neighboring roles.
\end{itemize}
All methods are evaluated under different problem sizes characterized by $S_{\max}$.

Table~\ref{tab:benchmark_results} compares the proposed methods with several others across different problem sizes. The ILP-based method is denoted as Batched ILP and the GRASP-based method as GRASP. We make several observations.
First, both the ILP-based method and the GRASP-based method consistently achieve high yield across all benchmarks and significantly outperform the greedy and random methods. For Benchmarks A and B, greedy and random methods produce yields close to zero. In contrast, the proposed methods maintain stable and high-quality results, demonstrating the effectiveness of the modeling and preprocessing techniques.
Second, the ILP-based method achieves solution quality very close to the monolithic ILP while improving scalability. On Benchmarks A and B, its median yield differs from the monolithic ILP by less than 1\%, while runtime decreases substantially as the problem size grows. For example, at $S_{\max}=1000$, it is several times faster than the monolithic ILP on Benchmark A. On Benchmark B, the monolithic ILP becomes intractable at larger sizes; for $S_{\max}=800$ and $S_{\max}=1000$, it fails to converge within the time limit, even after extended execution exceeding two weeks. In contrast, the ILP-based method remains tractable in all cases, showing that batching mitigates scalability limitations.
Third, the GRASP-based method further improves runtime while maintaining competitive solution quality. It is consistently faster than the ILP-based method, often by 2--5$\times$ on Benchmarks A and B, with yield only slightly lower, typically within a few percentage points. This indicates a favorable trade-off between quality and runtime, making it suitable for large-scale or time-constrained scenarios.
Finally, for Benchmarks D1 and D2, all methods achieve similarly high yields, suggesting weaker constraints. However, the ILP-based and GRASP-based methods are more effective than greedy and random methods, which still show noticeable degradation.

Overall, the ILP-based method provides near-optimal solutions with improved scalability, while the GRASP-based method offers higher efficiency with only minor loss in yield, enabling practical deployment across a wide range of problem sizes.

\subsection{Yield-Aware Cost Analysis of SoC Assembly}
\label{sec:cost}

\begin{table}[t]
\centering
\caption{Yield-aware cost savings of die matching methods at $S_{\max}=1000$. Random matching is used as the reference. Cost per good SoC is computed as $\text{total cost} / Y_{\text{mean}}$.}
\vspace{-0.5em}
\label{tab:cost_saving_s1000}
\footnotesize
\renewcommand{\arraystretch}{0.3}
\setlength{\tabcolsep}{2pt}

\begin{tabular}{|c|c|c|c|c|c|c|}
\hline
\textbf{ID} 
& \makecell{\textbf{Total cost} \\ \textbf{(\$)}}
& \textbf{Method} 
& \makecell{\textbf{$Y_{\text{mean}}$} \\ \textbf{(\%)}}
& \makecell{\textbf{Cost per} \\ \textbf{good SoC (\$)}}
& \makecell{\textbf{Cost saving} \\ \textbf{(\$)}}
& \makecell{\textbf{Saving} \\ \textbf{(\%)}} \\
\hline

\multirow{5}{*}{D1}
& \multirow{5}{*}{2,500}
& Random & 75.2 & 3,324 & 0     & 0.0 \\
& & Greedy               & 81.8 & 3,056 & 268   & 8.0 \\
& & \textbf{Batched ILP}          & 97.8 & 2,556 & 768   & 23.1 \\
& & \textbf{GRASP}                & 97.3 & 2,569 & 755   & 22.7 \\
& & Monolithic ILP       & 97.8 & 2,556 & 768   & 23.1 \\
\hline

\multirow{5}{*}{D2}
& \multirow{5}{*}{17,500}
& Random & 58.6 & 29,863 & 0      & 0.0 \\
& & Greedy               & 72.1 & 24,271 & 5,592  & 18.7 \\
& & \textbf{Batched ILP}          & 97.6 & 17,930 & 11,933 & 40.0 \\
& & \textbf{GRASP}                & 97.6 & 17,930 & 11,933 & 40.0 \\
& & Monolithic ILP       & 97.7 & 17,911 & 11,952 & 40.0 \\
\hline

\end{tabular}
\vspace{-1.5em}
\end{table}

To quantify the economic impact of chiplet matching in DT-MATCH, we convert yield into cost using a cost-per-good-SoC model based on the relationship $C_{\text{good}} = C_{\text{SoC}} / Y_{\text{mean}}$, where $C_{\text{SoC}}$ is the total cost and $Y_{\text{mean}}$ is the average assembly yield. Table~\ref{tab:cost_saving_s1000} reports results at $S_{\max}=1000$, with cost savings measured relative to random matching.
Yield improvements translate into cost reductions due to the inverse relationship between cost and yield. For D1, the ILP-based method increases $Y_{\text{mean}}$ from $75.2\%$ (Random) to $97.8\%$, reducing the cost per good SoC from \$3,324 to \$2,556, a $23.1\%$ reduction. The GRASP-based method achieves a similar result with $97.3\%$ yield and \$2,569 cost. For D2, the ILP-based method improves $Y_{\text{mean}}$ from $58.6\%$ to $97.6\%$, reducing cost from \$29,863 to \$17,930, corresponding to a $40.0\%$ reduction, with the GRASP-based method achieving the same result.
Greedy matching provides only moderate savings, reducing cost by $8.0\%$ for D1 and $18.7\%$ for D2, indicating that simple heuristics cannot fully exploit yield improvements. In contrast, the optimization-based methods consistently achieve the highest yield and lowest cost. The ILP-based method, the GRASP-based method, and the monolithic ILP produce nearly identical results on both benchmarks, indicating negligible yield differences at this scale.


\subsection{Ablation Study}
To validate the effectiveness of the enhancement techniques introduced in \cref{sec:framework}, we conduct an ablation study to quantify the impact of removing each component on runtime and the SoC yield. For the ILP-based approach, we examine the effect of excluding exact preprocessing. The rolling-horizon batching strategy is not further evaluated here, since it has been discussed in \cref{sec:main} and is equivalent to the monolithic ILP formulation. For the GRASP-based approach, we analyze the impact of removing exact preprocessing, as well as disabling reactive GRASP. \par

\begin{figure}[t]
    \centering
    \footnotesize
    \includegraphics[width=1\linewidth]{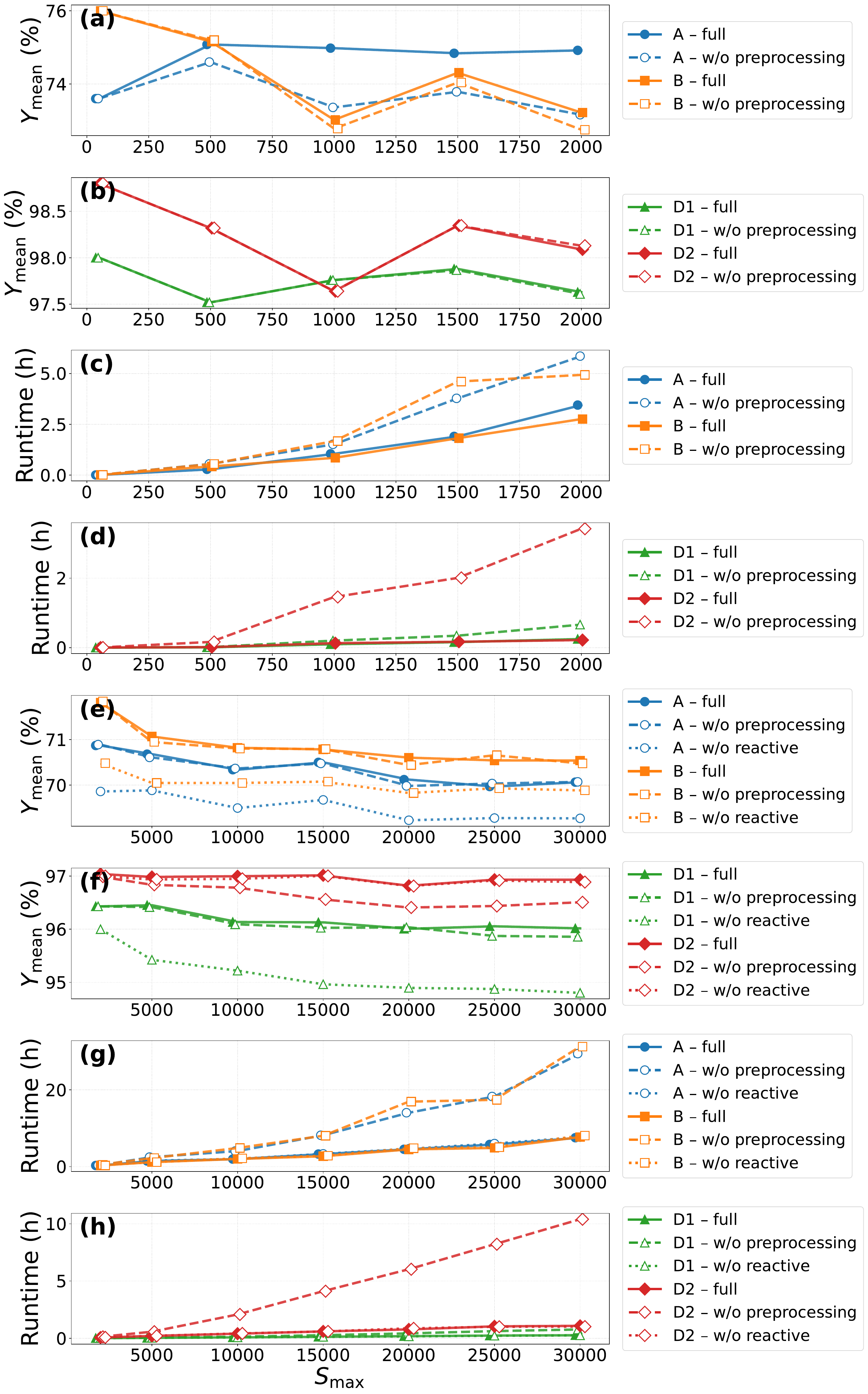}
    \vspace{-2em}
    \caption{Ablation study of preprocessing and reactive components.
(a) and (b) show $Y_{\text{mean}}$ of the ILP-based method with and without preprocessing on Benchmarks A and B, and D1 and D2, respectively.
(c) and (d) show the runtime of the ILP-based method under the same settings.
(e) and (f) show $Y_{\text{mean}}$ of the GRASP-based method under three configurations (full, without preprocessing, and without reactive mechanism) on Benchmarks A and B, and D1 and D2, respectively.
(g) and (h) show the runtime of the GRASP-based method under the same configurations.}
    \label{fig:ab_ilp}
    \vspace{-1.5em}
\end{figure}

\cref{fig:ab_ilp} presents an ablation study on the key components of the proposed methods. For the ILP-based method (\cref{fig:ab_ilp}(a)--(b)), removing preprocessing has little impact on yield. On Benchmark A, the difference is within about 2\%, and it is marginal on the others. A similar pattern is observed for the GRASP-based method (\cref{fig:ab_ilp}(e)--(f)). This behavior is expected because preprocessing groups instances with identical faulty-lane patterns into equivalence classes, so the method operates at the class level and makes decisions over aggregated representatives. Without preprocessing, the method instead operates directly on individual instances. Although this transformation is lossless in terms of feasible solutions, it alters the structure of the search space and the ordering of decisions, which leads to small variations in the final yield.
Preprocessing has a strong impact on runtime. As shown in \cref{fig:ab_ilp}(c)--(d) and \cref{fig:ab_ilp}(g)--(h), removing it significantly increases runtime, especially as $S_{\max}$ grows, confirming that preprocessing reduces problem size and improves scalability. The effect is less pronounced on Benchmark D1, which is less constrained.
For the GRASP-based method, disabling the reactive mechanism causes a small but consistent yield drop of about 1\% across all benchmarks, with negligible runtime change. This indicates that reactive GRASP improves solution quality with minimal overhead.
Therefore, preprocessing is critical for scalability with negligible impact on yield, while the reactive GRASP mechanism provides modest yield gains at almost no runtime cost.

\begin{figure}[t]
    \centering
    \footnotesize
    \includegraphics[width=0.8\linewidth]{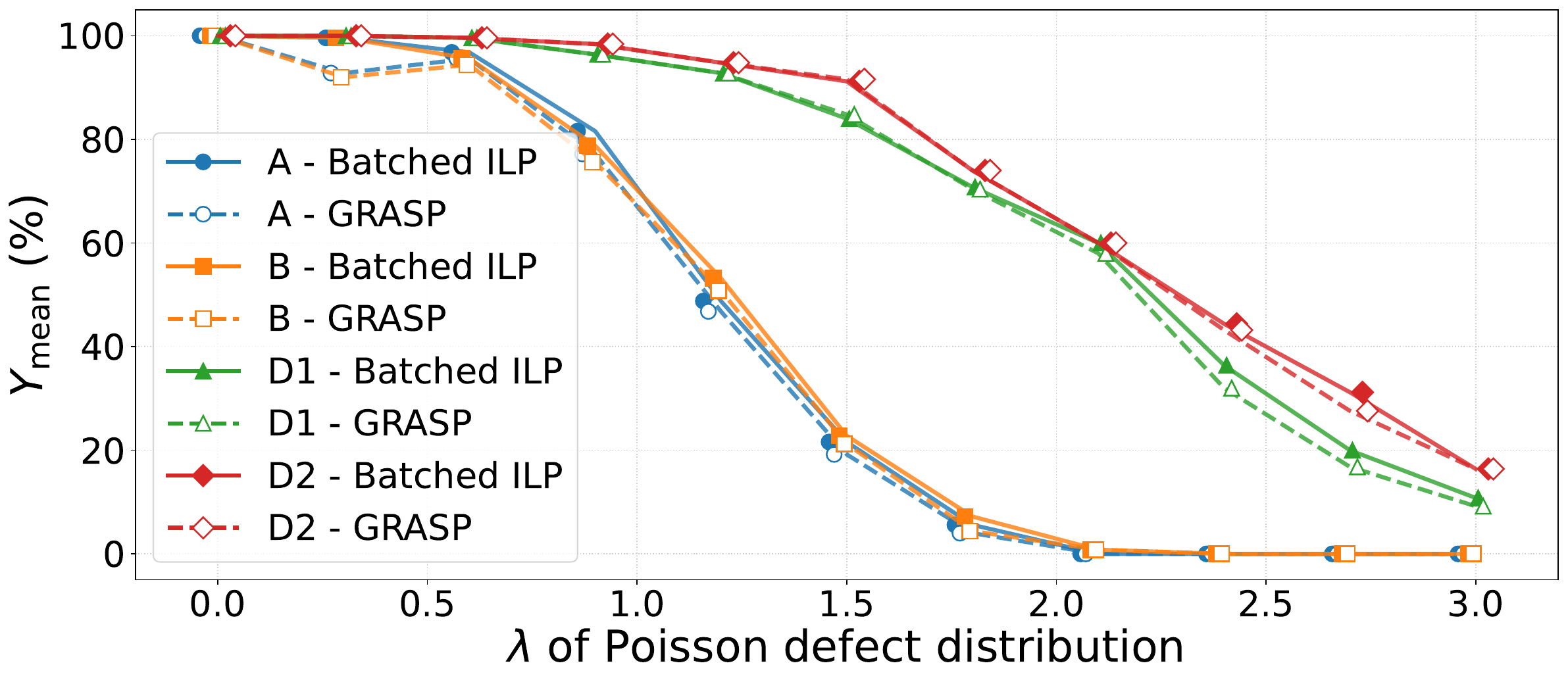}
    \vspace{-0.8em}
    \caption{Impact of the Poisson parameter $\lambda$ on yield across benchmarks for the ILP-based method (Batched ILP) and the GRASP-based method (GRASP) with $S_{\max}=250$.}
    \label{fig:ab_po}
    \vspace{-1.5em}
\end{figure}

We further evaluate the impact of the Poisson parameter $\lambda$ on $Y_{\text{mean}}$. As shown in \cref{fig:ab_po}, yield decreases as $\lambda$ increases across all benchmarks, since larger $\lambda$ leads to higher expected faulty-lane counts.
The sensitivity varies across benchmarks. For the more challenging benchmarks (A and B), yield drops rapidly and approaches zero when $\lambda \geq 2$, indicating that high defect rates severely limit feasible SoC assembly. In contrast, Benchmarks D1 and D2 are more robust, maintaining relatively high yield at moderate $\lambda$, although the decreasing trend remains clear.
Across all settings, the ILP-based method (Batched ILP) consistently achieves slightly higher yield than the GRASP-based method (GRASP), while both follow similar degradation trends as $\lambda$ increases. This indicates that the methods remain stable under varying defect levels.

\section{Conclusion}
\label{sec:conclusion}
We have presented DT-MATCH, a defect- and topology-aware chiplet matching framework that maximizes the number of functional SoCs. 
We formulate the post wafer-level test assembly problem, capturing role-specific chiplet types, lane-level defect maps, shared spare-lane budgets, and allowed chiplet orientations, and prove it to be NP-complete. We propose a unified lossless preprocessing pipeline including equivalence-class aggregation and role-level pruning to reduce problem size while preserving optimality. Building on this pipeline, we develop two complementary approaches: an ILP-based approach with rolling-horizon batching and periodic repair for high solution quality, and a GRASP-based approach with scarcity-guided construction for scalable performance. Experimental results on industry benchmarks show that the ILP-based approach achieves superior yield, while the GRASP-based approach provides significant runtime reduction with comparable solution quality, offering a practical trade-off for chiplet assembly optimization.

\section*{Acknowledgment}
The authors thank Arjun Chaudhuri from Arizona State University for his comments and suggestions on this work.

\balance

\bibliographystyle{IEEEtran}
\bibliography{reference_brief}
\end{document}